\documentclass[aps,prb,twocolumn,longbibliography,groupedaddress,showpacs,floatfix,superscriptaddress]{revtex4-1}
\usepackage{graphicx}  
\usepackage{dcolumn}   
\usepackage[english]{babel}
\usepackage{bm}        
\usepackage{amsmath, amsthm, amssymb}
\usepackage{bbold} 
\usepackage{mathrsfs}
\usepackage{array}
\usepackage[usenames]{color}
\usepackage{soul} 
\usepackage{ulem} 

\emergencystretch=\maxdimen
\newcommand{\calH}{\mathcal{H}}
\newcommand{\iv}{\mathbf{i}}
\newcommand{\jv}{\mathbf{j}}

\begin{document}

\normalem

\title{Coherence Temperature in the Diluted Periodic Anderson Model}
\author{N. C. Costa}
\affiliation{Instituto de F\'isica, Universidade Federal do Rio de
Janeiro Cx.P. 68.528, 21941-972 Rio de Janeiro RJ, Brazil}
\author{T. Mendes-Santos}
\affiliation{The Abdus Salam International Centre for Theoretical
Physics, strada Costiera 11, 34151 Trieste, Italy}
\author{T. Paiva}
\affiliation{Instituto de F\'isica, Universidade Federal do Rio de
Janeiro Cx.P. 68.528, 21941-972 Rio de Janeiro RJ, Brazil}
\author{N. J. Curro}
\affiliation{Department of Physics, University of California, Davis, CA 95616, USA}
\author{R. R. dos Santos} 
\affiliation{Instituto de F\'isica, Universidade Federal do Rio de
Janeiro Cx.P. 68.528, 21941-972 Rio de Janeiro RJ, Brazil}
\author{R. T. Scalettar}
\affiliation{Department of Physics, University of California, Davis, CA 95616,
USA}
\begin{abstract}
The Kondo and Periodic Anderson Model (PAM) are known to provide a
microscopic picture of many of the fundamental properties of heavy
fermion materials and, more generally, a variety of strong correlation
phenomena in $4f$ and $5f$ systems.  In this paper, we apply the
Determinant Quantum Monte Carlo (DQMC) method to include disorder in the
PAM, specifically the removal of a fraction $x$ of the localized
orbitals.  We determine the evolution of the coherence temperature
$T^*$, where the local moments and conduction electrons become entwined
in a heavy fermion fluid, with $x$ and with the hybridization $V$
between localized and conduction orbitals.  We recover several of the
principal observed trends in $T^*$ of doped heavy fermions, and also
show that, within this theoretical framework, the calculated Nuclear
Magnetic Resonance (NMR) relaxation rate tracks the experimentally
measured behavior in pure and doped CeCoIn$_5$.
Our results contribute to important issues in the 
interpretation of local probes of disordered, strongly
correlated systems.

\end{abstract}


\maketitle


\section{Introduction}
Materials poised at the cusp of magnetic to non-magnetic phase boundaries exhibit a myriad of complex properties.  
As systems ranging from cuprate superconductors\cite{scalapino95,keimer15} to heavy fermions\cite{tou95,coleman07,gegenwart08} and iron pnictides\cite{chubukov08,si16} are moved with pressure, chemical doping, or temperature away from a regime where magnetic order is dominant, an incredible variety of alternate patterns of spin, charge and pairing emerges.  
A description of the resulting competition has been an on-going challenge to the condensed matter community.\cite{dagotto05}

In order to address these phenomena,
Nuclear Magnetic Resonance (NMR) has been extensively used to explore local microscopic properties of correlated materials, providing great insight into their nature.\cite{kitaoka95,curro09,walstedt18}  
In particular, NMR experiments have determined the energy scale at which a heavy fermion state emerges, i.e.\,when $4f$-electrons become delocalized.  
This scale has been associated\cite{Curro04,Yang08,Yang12,Yang14,Yang15,Yang08b,Yang17} with a {\it coherence temperature,} $T^*$, whose signature appears, e.g.\,as an anomaly in Knight Shift (KS) measurements: While for normal metals the KS tracks the magnetic susceptibility, for most heavy fermion materials this tracking breaks down below a certain temperature, which is identified with $T^*$.\cite{Curro04,Yang08} 
The presence of several distinct contributions to the magnetic susceptibility in these materials, in particular the one from a singlet $d$-\!$f$ channel which delocalizes $4f$-electrons, leads to this anomaly, signalling the emergence of a heavy fermion state.  
Remarkably, when the KS anomaly is singled out by removing the high temperature contribution to the susceptibility, many heavy fermion materials exhibit a universal behavior for temperatures below $T^*$. \cite{Curro04,Yang08}
The coherence temperature is evident in multiple experimental probes,
including transport, thermodynamic, and tunnelling measurements,
but its microscopic origin, and its relation to the Kondo 
screening temperature remain open questions.\cite{Yang08b,wirth16,Yang17} 

Additional complexity is introduced by added chemical impurities, \cite{Miranda05,Kawasaki06,Seo14,wirth16,Chen18} so that treating the effects of disorder is essential to understand many of the properties of correlated electron materials.  
Randomness is central to the emergent physics since it acts to limit the growth of charge ordered regions.\cite{nie14}  
Likewise, dopant disorder can stabilize localized antiferromagnetic (AF) regions, explaining the persistence of AF even deep in the $d$-wave phase.\cite{andersen07}  
A  similar phenomenon occurs in heavy fermion materials where AF long range order is induced via Cd doping of CeCoIn$_5$.\cite{Seo14,pham06}  
Of particular interest is the crossover between Kondo screening in the single-impurity limit, and collective screening with intersite interactions among multiple sites in a lattice.  

A powerful approach to investigate these crossover regimes is to systematically replace the $f$-sites with non-magnetic atoms.  
This leads to inhomogeneities in the magnetic response, with some spatial regions favoring strong spin correlations, while in others a paramagnetic behavior is preferred.  
Thus, instead of having a single external parameter that globally tunes a system through a magnetic/non-magnetic boundary, one should also investigate how the physical quantities behave in the presence of internal, and highly inhomogeneous degrees of freedom.  
One expects that NMR quantities like $T^*$ and the spin-lattice relaxation rate to have a strong dependence with impurity doping (e.g.\,La substitution on Ce-based compounds) and
even acquire a distribution of values depending on the local environment 
of the nuclei.\cite{Nakatsuji02,Nakatsuji04,Ohishi09,Ragel09,Lawson18,MacLaughlin01,MacLaughlin04}
Indeed, NMR and scanning tunneling microscopy (STM) measurements on the cuprates have examined the links between charge order, superconductivuty, and pseudogap physics in the cuprates.

From a theoretical point of view,  the nature of these emergent phenomena
may be described by simplified models which take into account their most fundamental mechanisms, such as  the Periodic Anderson Model (PAM),\cite{gebhard97,fazekas99,Vekic95,hu17,Schafer18}, and the closely related Kondo Lattice model,\cite{doniach77,lacroix79,fazekas91,assaad99,costa17b} which consider weakly correlated `conduction' electrons hybridized with strongly correlated `localized' ones.
Tuning the strength of the hybridization in these models leads to a quantum phase transition (QPT), in which the ground state evolves from an antiferromagnetic (AF) ordering to a spin liquid state.  
Recent numerical work on the homogeneous PAM has captured the KS anomaly and provided $T^*$ by quantitatively characterizing the different orbital contributions to the global susceptibility.\cite{Jiang14,Jiang17}  
In the context of impurity doping,\footnote{Similar results were also found in a closely related spin system, as reported in Ref.\,\onlinecite{mendes-santos17}.} the PAM successfully describes the enhancement of AF correlations around doped impurities in CeCo(In$_{1-x}$Cd$_{x}$)$_{5}$,\cite{benali16,Wei17,costa18a}
and also provides evidence of a magnetic suppression when nonlocal hybridization terms are included\cite{Wu15,zhang19}, as in the case of CeCo(In$_{1-x}$Sn$_{x}$)$_{5}$.\cite{Sakai15,Chen18}

Here we study the combination of randomness and strong interactions with an exact numerical approach which allows for `real space imaging' of spin correlations.  
We investigate the behavior of the coherence temperature and NMR quantities in {\it chemically doped} heavy fermion materials, such as in Ce$_{1-x}$La$_{x}$CoIn$_{5}$, with quantum simulations which accurately incorporate sites with missing magnetic moments.
Our focus is on whether the calculated trends of these quantities resemble those from experimental NMR measurements\cite{Lawson18} on the evolution with impurity doping and external parameters.  
To this end, we extend previous\cite{Jiang14} Determinant Quantum Monte Carlo (DQMC) simulations to treat the {\it randomly diluted} PAM (dPAM),
as presented in the next Section.
In Sections \ref{Sec:Tstar} and \ref{Sec:1T1} we discuss our findings, from which
our key results are as follows: (1) the KS anomaly exhibits a universal scaling behavior below $T^{*}$, even in presence of disorder; (2) $T^{*}$ increases with hybridization $V$ (or pressure) and (3) linearly decreases with impurity concentration $x$.
Finally, (4) the NMR relaxation rate, $1/T_{1}$, exhibits a strongly inhomogeneous pattern throughout the lattice.
These demonstrate that several of the most fundamental conclusions of NMR experiments can be predicted, including the scaling behavior of $T^*$.
In Section \ref{Sec:conclusions} we summarize our main conclusions.

\section{Model and Method}\label{Sec:Model}

The Hamiltonian for the dPAM reads \footnote{An early attempt to investigate the
dPAM Hamiltonian was reported in Ref.\,\onlinecite{Costi88}.}
\begin{align}\label{eq:hamil}
	\calH=&-t\sum_{\langle\iv,\jv\rangle,\sigma}\left(c_{\iv\sigma}^\dagger 
c_{\jv\sigma}^{\phantom{\dagger}}+\mathrm{h.c.}\right)
	-\sum_{\iv,\sigma}V^{\phantom{\dagger}}_\iv\left(c_{\iv\sigma}^\dagger 
f_{\iv\sigma}^{\phantom{\dagger}}+\mathrm{h.c.}\right)\nonumber\\
&-\mu \sum_{\iv,\sigma,\alpha} n^{\alpha}_{\iv\sigma}
+ \sum_{\iv}U^{f}_{\iv}
\left(n^{f}_{\iv\uparrow}-\frac{1}{2}\right) 
\left(n^{f}_{\iv\downarrow}-\frac{1}{2}\right)
 \,\, ,
\end{align}	
where the sums run over a two-dimensional square lattice, with $\langle{\bf i,j} \rangle$ denoting nearest-neighbor sites, and $\alpha=c$ or $f$; the notation for the operators is standard.
The first term corresponds to the hopping of conduction electrons (the hopping integral, $t$, sets the energy scale), while the last  describes the Coulomb repulsion on localized $f$-orbitals.  
The hybridization between these two orbitals is modelled by a site-dependent hopping $V_\iv$

Here we consider {\it full orbital dilution}, in which we randomly set $U_{\bf i}=V_{\bf i}=0$ on a fraction $x$ of the sites.
Physically, this is equivalent to completely removing $f$-orbitals, similarly to the replacement of a magnetic $4f^{1}$ Ce atom by a $4f^{0}$ La one in CeCoIn$_{5}$, which locally suppresses both the moment on the $f$-orbital and the possibility of $c$-\!$f$ mixing (due to the distance of the La level from the Fermi energy).

The DQMC method\cite{blankenbecler81,hirsch85,white89,Loh90,dosSantos03b,troyer05} employed here to solve Eq.\,\ref{eq:hamil} is an unbiased technique commonly used to investigate Hubbard-like Hamiltonians: it maps a $d$-dimensional quantum system in a classical ($d$+1)-dimensional one, via the inclusion of an imaginary-time coordinate.   
Within this approach, one separates the one-body ($\hat {\mathcal K}$) and two-body ($\hat {\mathcal P}$) pieces in the partition function by using the Trotter-Suzuki decomposition, i.e.\,by defining $\beta = L_{\tau} \Delta \tau$, with $L_{\tau}$ being the number of imaginary-time slices, and $\Delta \tau$ the discretization grid.  
Then 
\begin{align}
	{\cal Z} &= {\rm Tr}\, e^{-\beta \hat {\mathcal H} } = {\rm Tr}\, \big[ \big(e^{-\Delta\tau ( \hat {\mathcal K} + \hat {\mathcal P})}\big)^{L_{\tau}} \big]\nonumber\\
	 &\approx {\rm Tr}\, \big[ e^{-\Delta\tau \hat
{\mathcal K}} e^{-\Delta\tau \hat {\mathcal P}} e^{-\Delta\tau \hat
{\mathcal K}} e^{-\Delta\tau \hat {\mathcal P}} \cdots \big], 
\end{align}
with an
error proportional to $(\Delta \tau)^2$.  
This is exact in the limit $\Delta \tau \to 0 $.  
The resulting partition function is rewritten in quadratic (single-body) form through a discrete Hubbard-Stratonovich transformation (HST) on the two-body terms, $e^{-\Delta \tau \hat{\mathcal P}}$.  
This HST introduces discrete auxiliary fields with components on each of the space and imaginary-time lattice coordinates, which are sampled by Monte Carlo techniques.
In this work we choose $t \Delta \tau=0.1$, so that the error from the Trotter-Suzuki decomposition is less than, or comparable to, that from the Monte Carlo sampling.
DQMC is able to measure a general set of single- and two-particle response functions, such as susceptibilities, which can be directly compared with experimental results.

Although numerically exact, DQMC is constrained by the infamous minus-sign problem,\cite{Loh90,troyer05} which restricts our analyses to the half-filling case, i.e.~when both $c$- and $f$-orbitals have $\langle n^{c,f}_{\mathbf{i},\sigma} \rangle=1/2$.
Determinant Quantum Monte Carlo is especially well matched to analyze the problem of disorder and the local structures which form around an impurity, since it is formulated in real space.  
Furthermore, many types of randomnesses such as local variations in hybridization, on-site repulsion, and site removal, do not affect particle-hole symmetry.
Therefore there is no sign problem at half-filling, regardless the presence of disorder (dilution) on the lattice.
This allows us to investigate the behavior of correlations in
all temperature scales.

\begin{figure*}[t]
\centering
\includegraphics[scale=0.62]{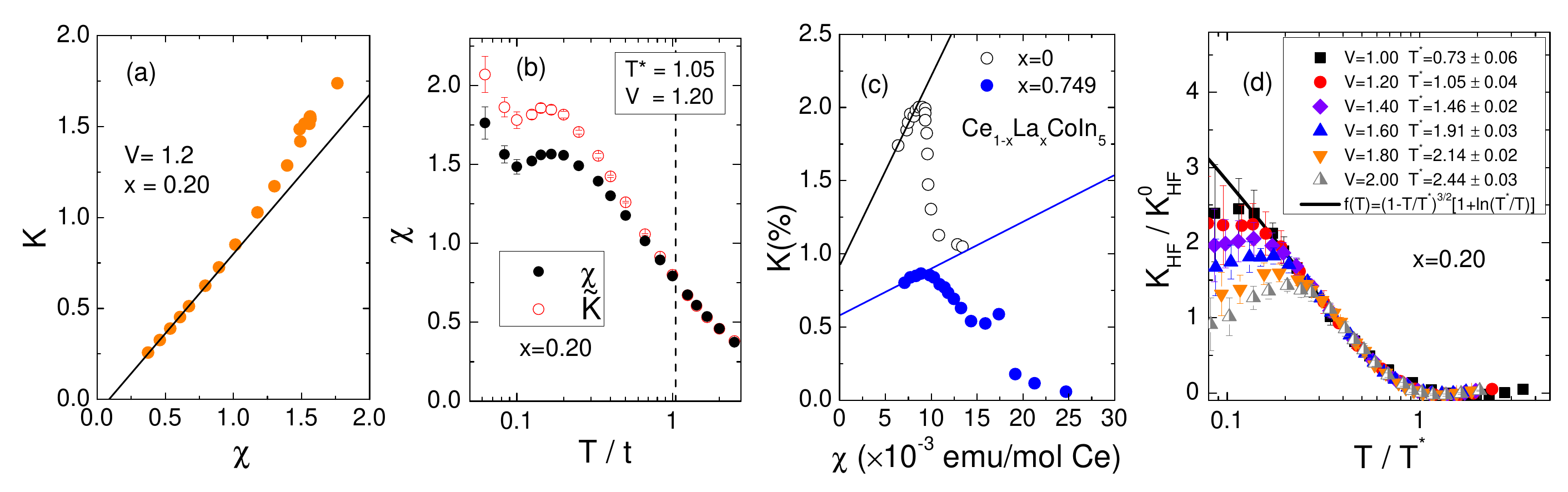} %
\caption{(Color online) (a) Knight shift as a function of total susceptibility
$\chi$. (b) $\chi$ and the renormalized Knight shift
$\tilde K$ as a function of temperature, for $V/t=1.2$ and $x=0.20$.
The vertical dashed line defines $T^{*}/t = 1.05$ (see text).
(c) Experimental NMR results for Ce$_{1-x}$La$_{x}$CoIn$_{5}$ reproduced
from Ref.\,\onlinecite{Lawson18}.  (d) data collapse of DQMC results for
the KS anomaly, $K_{HF}$, for $T \lesssim T^*$.
Here, and in all subsequent figures, when not shown, error bars are smaller than
symbol size.}
\label{fig:Fig1} 
\end{figure*}

To connect with NMR measurements, the central
quantities of interest are magnetic susceptibilities, from which the
Knight Shift and spin-lattice relaxation rate are obtained; see below.
Due to the presence of two orbitals, the total spin on a given site
$\iv$ is $\mathbf{S}_{\mathbf{i}} = \mathbf{S}^{c}_{\mathbf{i}} +
\epsilon_{\iv} \mathbf{S}^{f}_{\mathbf{i}}$, with $\epsilon_{\iv} \equiv
1 $ at sites containing $f$-orbitals, and 0 otherwise.  Thus, the total
magnetic susceptibility is given by
\begin{equation}
	\chi = \chi_{cc}+2\chi_{cf}+\chi_{ff},
\end{equation}
where
\begin{equation}
\chi_{\alpha\alpha^\prime}=\frac{1}{N_s}\sum_{\iv\jv} Q_{\iv \jv}^{\alpha \alpha^\prime} 
\int_0^\beta \mathrm{d}\tau\, 
	\langle\mathbf{S}^\alpha_{\iv}(\tau) \cdot
\mathbf{S}_{\jv}^{\alpha^\prime}(0)\rangle .
\label{eq:suscab}
\end{equation}
Here $\mathbf{S}^{\alpha}_{\mathbf{i}}(\tau) = 
e^{\tau {\cal H}} \, \mathbf{S}^{\alpha}_{\mathbf{i}}(0) 
\, e^{-\tau {\cal H}}$, with $\alpha,\alpha^\prime=c$ or $f$, and 
$Q_{\iv \jv}^{\alpha \alpha^\prime} = \big[ \big( \delta_{\alpha, c} +
\epsilon_{\iv} \epsilon_{\jv} \delta_{\alpha, f} \big) \delta_{\alpha, \alpha^\prime}
+ \big( \epsilon_{\iv} \delta_{\alpha, f} \delta_{\alpha^\prime, c} +
\epsilon_{\jv} \delta_{\alpha, c} \delta_{\alpha^\prime, f} \big) \big] $;
the number of lattice sites is $N_s=L \times L$.
Similarly, the Knight shift is
\begin{align}
	 K   &= A\chi_{cc}+(A+B)\chi_{cf}+B\chi_{ff}+K_0,
\label{eq:K}
\end{align}
where $A$ ($B$) corresponds to the hyperfine coupling between the
nuclear spin of In(1) atoms and conduction (localized) electrons.  $K_0$
is a temperature-independent term arising from orbital and diamagnetic
contributions to $K$, which we set to zero.
Recall that the Knight Shift is a local quantity, which depends on the
distribution of nearest-neighbor site (Ce or La) moments to the central
In(1) atom.\cite{Lawson18}  
Thus, our~data correspond to the average $K$
of representative sites that couple to both $\mathbf{S}^{c}$ and
$\mathbf{S}^{f}$ spins.  
Since the hyperfine couplings are generally
different,\cite{Curro01,Curro04} and strongly material dependent, we
follow a previous study \cite{Jiang14} and take $A/B = 0.3$; general
trends are not sensitive to the precise choice of $A/B$.\cite{Jiang14}
Our simulations capture qualitative features of $\chi$, $K$ and
$1/T_1$, but not material-specific details.  In what follows, our DQMC
data are averages over 20-30 different disorder configurations on a
$10\times10$ square lattice, and $U^f/\,t=4$.  
Most of our results were
obtained for $V\! \geq\! 1$, corresponding to the singlet region for the
clean PAM.\cite{hu17,Schafer18}

\section{The Coherence Temperature}\label{Sec:Tstar}


At high temperatures, localized electrons are weakly coupled to conduction bands, so the contribution of $\chi_{cc}$ and $\chi_{cf}$ may be disregarded.  
As a result, the Knight Shift [Eq.\eqref{eq:K}] tracks the {\it localized} electron susceptibility and, under the same assumptions, the total susceptibility $\chi$ as well.
Following the procedure adopted in analyses of the experiments, we perform a linear fit to our DQMC Knight Shift data as a function of the susceptibility in the high temperature region,
i.e.~$K=B_\mathrm{eff}\chi+K_{0,\mathrm{eff}}$ [see Fig.\,\ref{fig:Fig1}\,(a)]. 
Next, we define the renormalized KS, 
\begin{equation}
	\widetilde{K} \equiv (K-K_{0,\mathrm{eff}})/B_\mathrm{eff},
\label{eq:KHF}
\end{equation}
which is equal to $\chi$ at high temperatures.  
This equality holds as
long as the $c$-\!$f$ singlet channel is small.  However, since $(A+B)/B \neq 2$,
$\tilde{K}$ fails to track $\chi$ when $\chi_{cf}$ becomes relevant: 
the associated energy scale is $T^*$; see Fig.\,\ref{fig:Fig1}\,(b). 
In this way the Knight Shift, which detects the
contribution of a $c$-\!$f$ channel (hence the presence of delocalized
4$f$-electrons), provides an important tool to investigate the emergence of a
heavy fermion state and its temperature scale.
The KS anomaly persists even in strongly diluted materials, as displayed in Fig.\,\ref{fig:Fig1}\,(c), for 
the heavy fermion Ce$_{1-x}$La$_{x}$CoIn$_{5}$, with $x\approx0.75$.

It is worth emphasizing that the continued appearance of the coherence temperature in the presence of disorder at a value similar to that of the pure system\cite{Jiang14,Raczkowski19} is a non-trivial observation.  
Indeed, electrons in unpaired orbitals (which survive dilution) are known to give large contributions to the susceptibility, regardless of whether they are conducting or localized.\cite{charlebois15,mendes-santos17,costa18a} 
This could, in principle, significantly affect the assumptions under which $\tilde K$ would track $\chi$, hence the value of $T^{*}$ as well.  
As we shall
see, these effects are more relevant at very low temperatures, due to the
possibility of long-range order setting in the ground state.  
The relatively weak dependence of the NMR quantities with dilution, 
as presented in Figs.\,\ref{fig:Fig1}\,(a)-(c), is an important step towards
a \textit{global} understanding of $T^*$ in diluted systems.

\begin{figure}[t]
\includegraphics[scale=0.25]{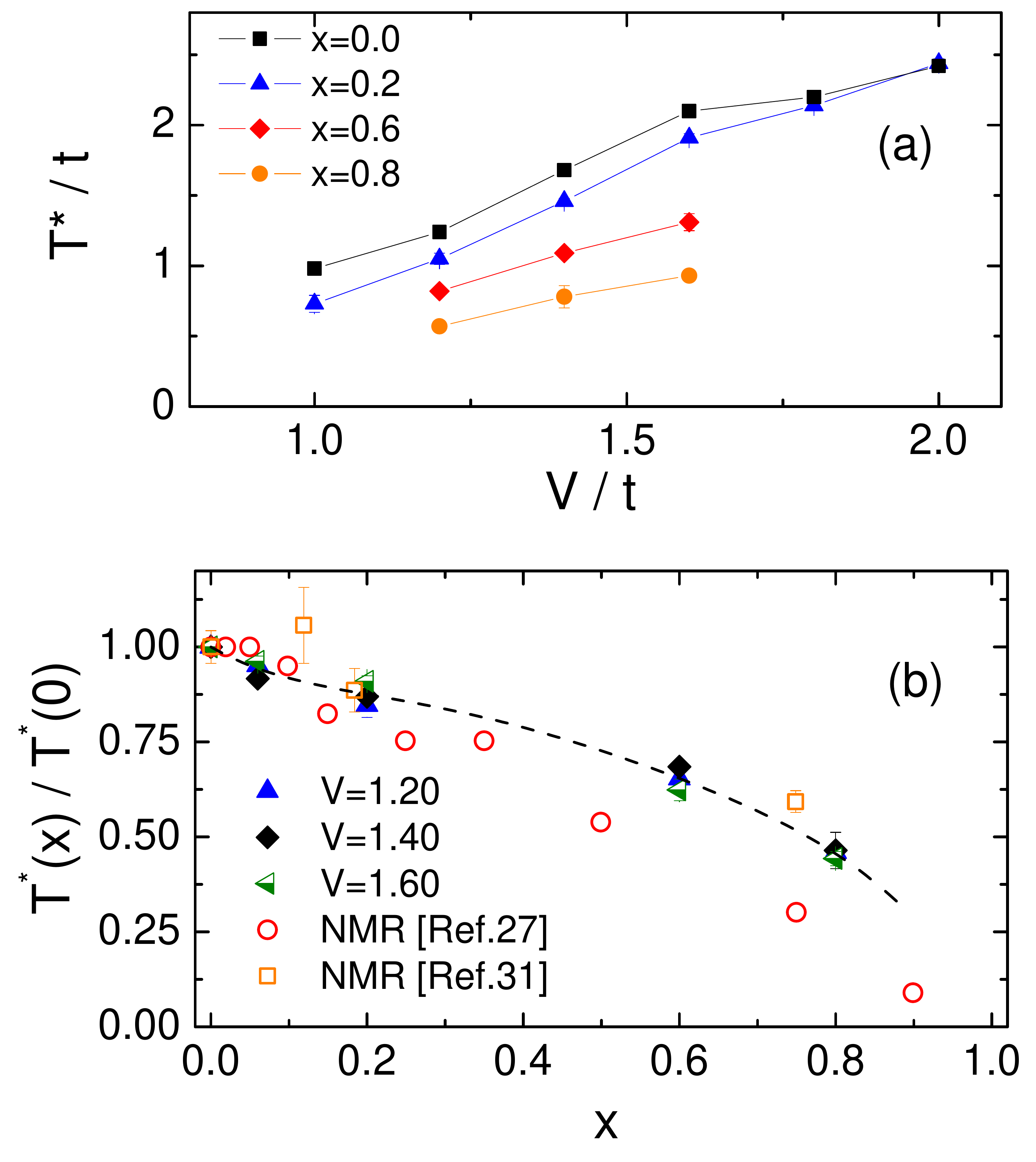} %
\caption{(Color online) (a) The coherence temperature $T_*$ as a function of $V$ for different dilution fractions $x$.
(b) Dependence of $T^*$ on $f$-orbital
dilution $x$ in the PAM.  Data for different $c$-$f$ hybridization $V$
are normalized to their clean ($x=0$) system value.  Experimental NMR
results on Ce$_{1-x}$La$_x$CoIn$_5$ are reproduced from
Ref.\,\onlinecite{Nakatsuji02} (open red circles), and
Ref.\,\onlinecite{Lawson18} (open orange squares).
The black dashed line is a guide to the eye for DQMC data.
The DQMC results are in excellent qualitative 
(and, indeed, almost quantitative) agreement
with experiment.}
\label{fig:Fig2} 
\end{figure}

Within a two-fluid model,\cite{Nakatsuji04,Curro04,Yang08,Shirer12,Yang16} one singles out the `heavy fermion fluid' contribution to the KS by subtracting its `normal' (high temperature) contribution, i.e.
\begin{equation}
	K_\mathrm{HF}\equiv B_\mathrm{eff} (\widetilde{K}-\chi).
\end{equation}	
Remarkably, experimental results suggest a universal behavior of $K_\mathrm{HF}$ for many different heavy fermion materials,  
\begin{equation}\label{eq:scaling_func}
	\frac{K_\mathrm{HF}(T)}{K^0_\mathrm{HF}}  
		= f(T) \equiv \, \left(1-T/T^*\right)^{3/2}\left[1+\ln (T^*/T) \right],
\end{equation}
where $K^0_\mathrm{HF}$ and $T^{*}$ depend on the specific material, and on external parameters.  
We use this phenomenological scaling form for a more accurate estimate of $T^{*}$ through the collapse of our DQMC data, as shown in Fig.\,\ref{fig:Fig1}(d).

The behavior of the KS in Fig.\,\ref{fig:Fig1}, in particular its scaling behavior [Fig.\,\ref{fig:Fig1}(d)], provides a robust evidence that DQMC simulations qualitatively reproduce trends observed experimentally, even in the presence of disorder.  
We now turn our attention to the dependence of $T^*$ with external parameters, such as the hybridization, $V$, which is tuned in experiments by applying pressure.  
Figure \ref{fig:Fig2}\,(a) displays the behavior of $T^*$ as a function of $V$ for different impurity concentrations, $x$.  
Regardless of the level of disorder, the coherence temperature increases monotonically with $V$.  
This reproduces fundamental features of NMR measurements (e.g., for CeRhIn$_5\,\,$\cite{Lin15}): larger hybridization increases the probability of a hopping from $f$-orbitals to conduction ones, which in turn increases the energy scale ($\sim V^2/U^f$). 

The effect of dilution on $T^*$ is already apparent in Fig.\,\ref{fig:Fig2}\,(a): although the clean and
disordered cases share the same qualitative trend, the value of $T^{*}$ decreases with $x$.  
This reduction in the coherence temperature with $f$-orbital dilution
reflects a crossover between dense and diluted Kondo regimes;
that is, the material goes from a heavy-fermion state at small $x$ to a single-impurity Kondo regime at $x \approx 1-\epsilon$, with $\epsilon \ll 1$.
To further emphasize this crossover, Fig.\,\ref{fig:Fig2}\,(b) displays $T^{*}$ as function of dilution, for different values of hybridization.  
Notice that $T^{*}$ has a (roughly) linear dependence with $x$, with $T^{*} \neq 0$ even at strong dilution.  
Our DQMC predictions are in good agreement with recent NMR results for Ce$_{1-x}$La$_{x}$CoIn$_{5}$, as shown in Fig.\,\ref{fig:Fig2}\,(b); see Ref.~\onlinecite{Lawson18}.  
Data from early attempts to measure $T^{*}$ in Ce$_{1-x}$La$_{x}$CoIn$_{5}$ (see, e.g.\,Ref.\,\onlinecite{Nakatsuji02}) are also included in Fig.\,\ref{fig:Fig2}\,(b): they also display a monotonic decrease of the coherence temperature with La doping.

\section{Relaxation Time}\label{Sec:1T1}

The NMR relaxation rate is defined as (see, e.g.\,Ref.\,\onlinecite{curro09})
\begin{equation}
T_{1}^{-1} = \gamma^2 k_{B} T \lim_{\omega \to 0} \sum_{\mathbf{q}} A^{2}(\mathbf{q})\, \frac{\chi^{''}(\mathbf{q},\gamma) }{\hbar \omega},
\end{equation}
where $ A^{2}(\mathbf{q}) $ is the square of the Fourier transform of
the hyperfine interaction, and $\gamma$ is the gyromagnetic ratio.  The
latter is related to the nuclear magnetic moment by $\gamma \hbar = g
\mu_{N} \sqrt{I(I+1)}$, with $\mu_{N}$ being the nuclear magneton, $g$
the nuclear $g$-factor, and $I$ the nuclear spin.
$T_{1}^{-1}$ quantifies a characteristic time in which a component of the nuclear spin (of a given site) reaches equilibrium after an external perturbation (magnetic field pulse).
It is a dynamical (real frequency) quantity whose numerical evaluation usually requires an analytic continuation of the imaginary-time DQMC data.
Instead, we use an approximation to this procedure,\cite{Randeria92}
\begin{align}
\label{eq:spin-relaxation2}
\frac{1}{T_{1}} = 
\frac{1}{\pi^2 T}
\sum_{\mathbf{i}} 
\bigg\langle 
S_{\mathbf{i}}(\tau=\beta/2) S_{\mathbf{i}}(0) 
\bigg\rangle \, .
\end{align} 

\begin{figure}[t]
\includegraphics[scale=0.26]{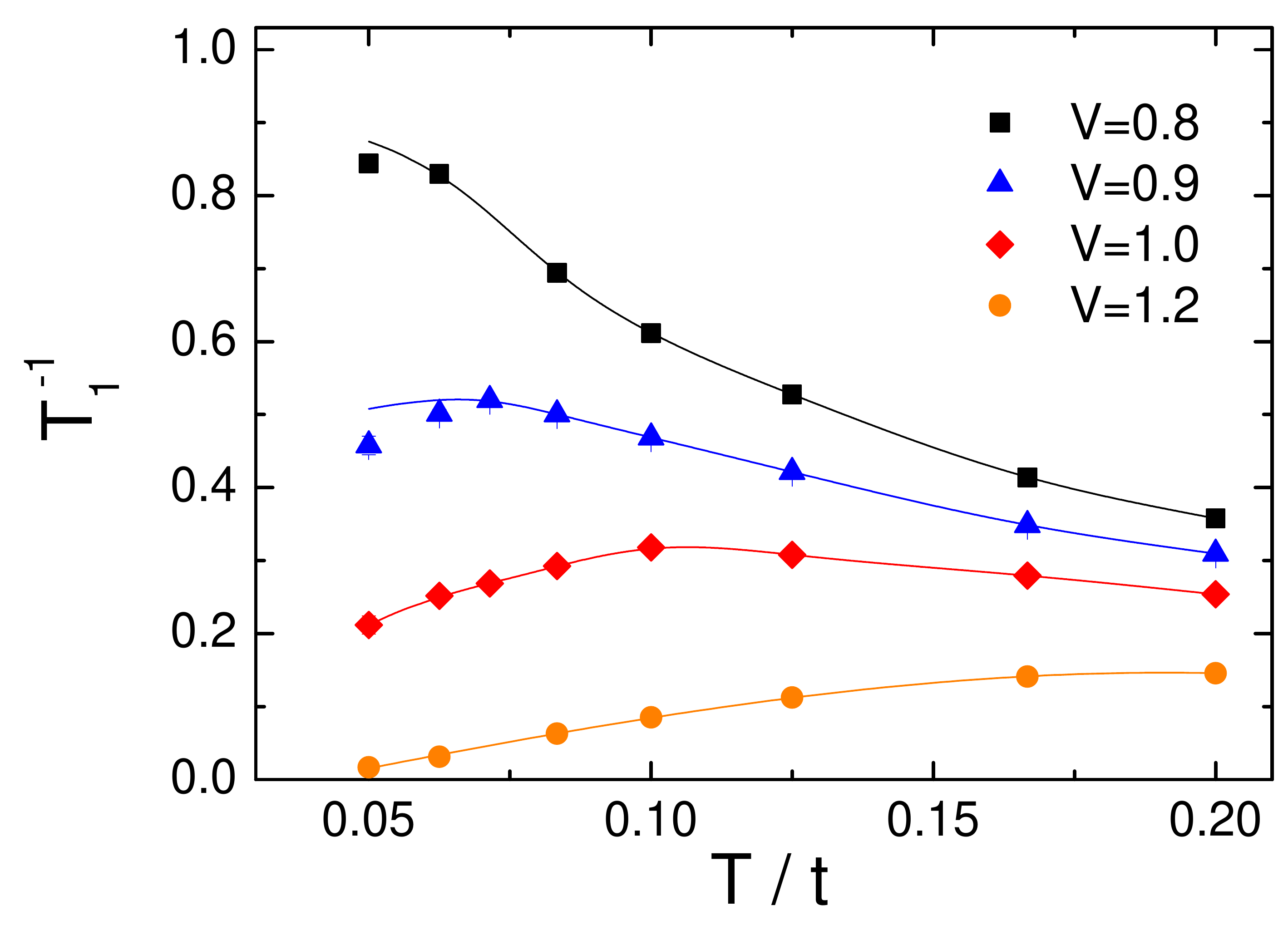} %
\caption{(Color online) Spin-lattice relaxation rate of the clean PAM as
a function of temperature, for different hybridizations $V$.
Solid lines are guides to the eye.}
\label{fig:T1_1} 
\end{figure}

As a benchmark for our results for the diluted case, we first examine the behavior of
$T^{-1}_{1}$ for the clean ($x=0$) system.
Previous DQMC studies\cite{Vekic95,hu17} of the PAM have provided evidence of a QPT from an antiferromagnetically (AFM) ordered ground state to a spin liquid phase at $V_{c} \approx 1.0$.  
Then, one expects that the behavior of $T_{1}^{-1}$ for decreasing temperatures should reflect the properties of these different ground states.\cite{mendes-santos17}
Figure \ref{fig:T1_1} displays the behavior of the relaxation rate as a
function of temperature for different values of $V$.  Here we show the
results from extrapolating data for lattice sizes $L=8$, 10, and 12 to
$L\to\infty$.  Within the AFM phase, $V/t =0.8$ or $0.9$, $T_{1}^{-1}$
approaches a finite nonzero value as $T \to 0$, consistent with the
absence of a spin gap, i.e. \!the presence of spin-wave excitations.  On
the other hand, for larger $V$, $T_{1}^{-1}$ decreases monotonically
when $T$ is lowered, reflecting a spin gapped (spin liquid) ground
state.  Notice that the change in behavior of $T_{1}^{-1}$ occurs around
$V/t \sim 1.0$, in line with the $V_{c}$ reported in
Ref.\,\onlinecite{hu17}.

\begin{figure}[t]
\includegraphics[scale=0.29]{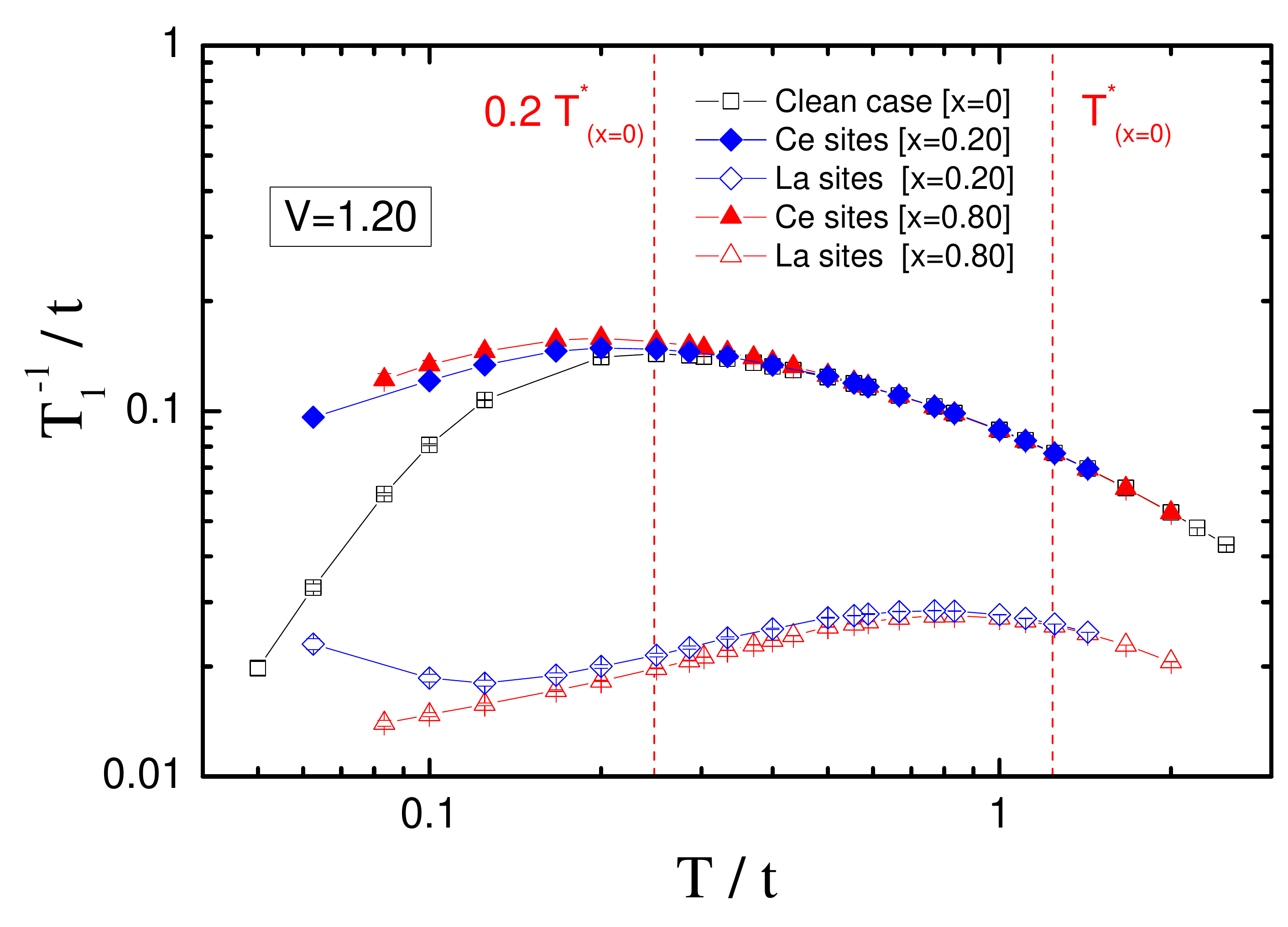} %
\caption{(Color online) Local contributions to the spin-lattice
relaxation rate for $x=0.20$ and 0.80, as comparison with the clean case
($x=0$).
Vertical red dashed lines correspond to $T^{*}$ and $0.2T^{*}$ for $x=0$, and $V/t=1.20$. 
}
\label{fig:T1_2} 
\end{figure}

Turning to the disordered case, the lack of translational symmetry requires the analysis of
\textit{local} contributions to $T_{1}^{-1}$, by considering two
species of sites: (i) Ce sites, those with an active $f$-orbital, and
(ii) La sites, those which had their $f$-orbitals removed.
Accordingly, we define $T_{1}^{-1}$ for Ce and La as the average over their
individual contributions, i.e.~we average over the available sites of
each type, and subsequently we average over disorder configurations.
Figure \ref{fig:T1_2} displays the behavior of the local $T_{1}^{-1}$ for fixed $V/t=1.2$ and for different concentrations.
For reasons which will become apparent below, we separate the discussion of
Fig.\,\ref{fig:T1_2} into two regimes: intermediate temperatures, $T\sim
T^*$, and low temperatures, $T\ll T^*$, when properties reflect the
dominant correlations in the ground state.  In the intermediate
temperature range, we note that data for the spin relaxation rate on Ce
sites for the clean case and for both dilution cases ($x=0.20$ and 0.80)
are almost indistinguishable; for the La sites, data for these same
concentrations are also nearly identical, though much smaller than those
for the Ce sites.  
When compared with the experimental results in Fig.\,10 of Ref.\,\onlinecite{Lawson18}, we see that the same data grouping occurs, and that the decrease of $T_{1}^{-1}$ as the temperature decreases (below the broad maxima) is also present; the difference in magnitude between data for Ce and La sites is also noticeable.  
These features therefore provide evidence that the
$T_{1}^{-1}$ distribution is quite inhomogeneous throughout the lattice,
with Ce sites behaving as in the clean case even for strong dilution.
A possible explanation for this inhomogeneity may be a local nature of
singlet formation, i.e.\,singlets have a short correlation length.

In the low-temperature regime, the strong attenuation observed in our
DQMC results for the pure case is due to the spin gapped ground state.  
For the diluted systems, however,
our $T_{1}^{-1}$ data on Ce sites seem to converge to finite values as
$T$ decreases, consistent with gapless behavior due to either enhanced
magnetic correlations or metallic (Pauli-like) behavior, depending on
the degree of dilution.
It is worth noting that the $T_{1}^{-1}$ for $x=0.20$ and $x=0.80$ have similar behavior, despite the large difference in the disorder strength.
In fact, previous theoretical works\cite{Kaul07,Watanabe10} have suggested that the dense Kondo regime occurs
just for $n_{c}\approx n_{f}$, while the diluted Kondo regime is established for a wide region of $n_{c} < n_{f}$, which is in line with our findings here.
The difference between these two dilutions occurs only for La sites at very low temperatures.
The data for La sites when $x=0.20$ show that
$T_{1}^{-1}$ increases with decreasing temperatures for $T/t\lesssim
0.1$, corresponding to an enhancement of magnetic correlations on these
sites, a behavior also found for the regularly depleted PAM.\cite{costa18a}
We note that the half-filling of our model for dilution may impose a bias towards an AFM ground state, since conduction sites with removed partners are unable to form singlets.\cite{mendes-santos17}

\section{Conclusions}\label{Sec:conclusions}

In summary, we have presented results for the magnetic susceptibility,
Knight Shift, and NMR relaxation rate computed using DQMC simulations
for the diluted Periodic Anderson Model.  We showed that even in the presence of
disorder, the Knight Shift anomaly displays a behavior with a
phenomenological universal function shared with the clean case.
We have also
obtained the coherence temperature, $T^{*}$, and its dependence on
$c$-\!$f$ hybridization, $V$, and with the dilution fraction $x$.  
We have found that $T^{*}$ is a linearly decreasing function of $x$, reproducing
a crucial feature of the experimental results for La-doped CeCoIn$_5$.
Finally, we have also discussed the spin-lattice relaxation rate, which
is distributed inhomogeneously throughout the lattice.
The qualitative agreement of our results
with experimental NMR measurements for Ce$_{1-x}$La$_{x}$CoIn$_5$ suggests
DQMC is a powerful theoretical
tool to model accurately the nature of spin correlations in disordered heavy
fermion materials.

Although we have emphasized the use of DQMC within the context of condensed matter materials, our work also has important implications for ``quantum gas microscopes" and their use to explore ultracold trapped atoms.\cite{cheuk15,kuhr16,ott16}
Like the NMR measurements described
here, quantum gas microscopy allows the resolution of single atoms,
doubly occupied sites, and (local) magnetic order.  A central focus is
on nonequilibrium properties directly connected to the relaxation times
studied here.

\begin{acknowledgments}
NCC, TP, and RRdS were supported by the Brazilian Agencies CAPES, CNPq,
and FAPERJ.  NJC was supported by NSF, grant number DMR-1005393 and
DMR-1807889, and RTS by Department of Energy, grant number DE-SC0014671.
\end{acknowledgments}

\bibliography{bib_rrds}
\end{document}